\documentclass[lettersize,journal]{IEEEtran}
\usepackage{amsmath,amsfonts}
\usepackage{algorithmic}
\usepackage{algorithm}
\usepackage{array}
\usepackage[caption=false,font=normalsize,labelfont=sf,textfont=sf]{subfig}
\usepackage{textcomp}
\usepackage{stfloats}
\usepackage{url}
\usepackage{verbatim}
\usepackage{graphicx}
\usepackage{cite}
\usepackage{pgfplots}
\usepackage{siunitx}
\usepackage{enumitem}

\usepackage[cmintegrals,cmbraces,varbb]{newtxmath} 
\usepackage[cal=cm,bb=ams]{mathalpha} 
\renewcommand{\ttfamily}{\fontencoding{T1}\fontfamily{lmtt}\selectfont}

\usepackage[hidelinks]{hyperref}
\usepackage[nameinlink]{cleveref}
\crefname{equation}{}{}
\crefname{algorithm}{Algorithm}{Algorithms}
\crefname{subsection}{Section}{Sections}
\crefname{section}{Section}{Sections}
\crefname{figure}{Figure}{Figures}

\newcommand{\R}{\mathbb{R}}
\newcommand{\Z}{\mathbb{Z}}
\DeclareMathOperator{\mpc}{MPC}

\usepackage{booktabs}

\usepackage{subcaption}
\def\tablename{Table}

\usepackage{tikz}
\usetikzlibrary{arrows.meta}

\definecolor{red1}{HTML}{E3120B} 
\definecolor{red2}{HTML}{CC100A} 
\definecolor{red3}{HTML}{F6423C} 
\definecolor{red4}{HTML}{FEE7E7} 

\definecolor{backgroundred1}{HTML}{E1D0D0} 
\definecolor{backgroundred2}{HTML}{EBE0E0} 
\definecolor{backgroundred3}{HTML}{F5F0EF} 

\definecolor{blue1}{HTML}{141F52} 
\definecolor{blue2}{HTML}{1F2E7A} 
\definecolor{blue3}{HTML}{2E45B8} 
\definecolor{blue4}{HTML}{475ED1} 
\definecolor{blue5}{HTML}{D6DBF5} 
\definecolor{blue6}{HTML}{EBEDFA} 

\definecolor{backgroundblue1}{HTML}{D0D3E1} 
\definecolor{backgroundblue2}{HTML}{E0E2EB} 
\definecolor{backgroundblue3}{HTML}{EFF0F5} 

\definecolor{green1}{HTML}{4C9C16} 
\definecolor{green2}{HTML}{62C91D} 
\definecolor{green3}{HTML}{7BE236} 
\definecolor{green4}{HTML}{E2F9D2} 
\definecolor{green5}{HTML}{F0FCE9} 

\definecolor{yellow1}{HTML}{F9C31F} 
\definecolor{yellow2}{HTML}{FBD051} 
\definecolor{yellow3}{HTML}{FCDE83} 
\definecolor{yellow4}{HTML}{FEF2CD} 
\definecolor{yellow5}{HTML}{FEF8E6} 

\definecolor{orange1}{HTML}{F97A1F} 
\definecolor{orange2}{HTML}{FB9851} 
\definecolor{orange3}{HTML}{FCB583} 
\definecolor{orange4}{HTML}{FEE1CD} 
\definecolor{orange5}{HTML}{FEF0E6} 

\definecolor{gray1}{HTML}{0D0D0D} 
\definecolor{gray2}{HTML}{1A1A1A} 
\definecolor{gray3}{HTML}{333333} 
\definecolor{gray4}{HTML}{595959} 
\definecolor{gray5}{HTML}{666666} 
\definecolor{gray6}{HTML}{B3B3B3} 
\definecolor{gray7}{HTML}{D9D9D9} 
\definecolor{gray8}{HTML}{F2F2F2} 
\definecolor{gray9}{HTML}{FFFFFF} 

\colorlet{nominal_hull_color}{blue1}
\colorlet{nominal_hull_color_fill}{blue1}
\colorlet{adapted_hull_color}{red1}
\colorlet{adapted_hull_color_fill}{red1}
\colorlet{origin_color}{gray1}
\colorlet{x0_color}{gray1}
\colorlet{zero_error_value}{backgroundred3}
\colorlet{full_error_value}{red1}
\colorlet{zero_error_input}{backgroundblue3}
\colorlet{full_error_input}{blue1}
\colorlet{axis_left_color}{blue1}
\colorlet{axis_right_color}{orange1}

\pgfplotsset{
    axis line style = {color=gray1,semithick},
    tick style = {color=gray1,semithick},
    tick label style = {color=gray1},
    xlabel style = {color=gray1,font=\small},
    every axis y label/.append style = {color=gray1,font=\small},
    legend style = {
        draw=gray1,
        style={text=gray1},
        semithick,
        font=\small
    },
    grid style = {gray7},
    tick label style={font=\scriptsize},
    grid=both,
    legend cell align=left,
    title style = {font=\small},
    colorbar style={
        tick label style={font=\scriptsize},
        label style={font=\small},
    }
}

\pgfplotsset{solid/.style={thick}}
\pgfplotsset{dashed/.style={dash pattern=on 5pt off 2pt}}
\pgfplotsset{dashdot/.style={dash pattern=on 5.25pt off 2pt on 2pt off 2pt}}
\pgfplotsset{mark_square/.style={mark=square*, mark size=1pt}}
\pgfplotsset{mark_halfsquare/.style={mark=halfsquare*, mark size=1pt}}

\pgfplotsset{benchmark/.style={solid,ultra thick,gray6}}
\pgfplotsset{no_adapt/.style={solid,mark=square*,mark options={solid},mark size=1pt,blue1}}
\pgfplotsset{adapt/.style={solid,mark=*,mark options={solid},mark size=1.2pt,yellow1}}
\pgfplotsset{adapt_k/.style={solid,mark=square*,mark options={solid,rotate=45},mark size=1pt,red1}}
\pgfplotsset{no_adapt_success/.style={solid,ultra thick,blue1!30}}
\pgfplotsset{adapt_success/.style={solid,ultra thick,yellow1!30}}
\pgfplotsset{adapt_k_success/.style={solid,ultra thick,red1!30}}

\newif\ifmainfile
\mainfilefalse

\begin{document}

\title{\textbf{Model-Agnostic Meta Learning for Differentiable MPC}}

\author{
  Salma Elfeki\textsuperscript{1}, %
  Riccardo Zuliani\textsuperscript{1}, %
  Niklas Schmid\textsuperscript{1}, %
  Efe C. Balta\textsuperscript{1,2}, %
  and John Lygeros\textsuperscript{1}%
  \thanks{
    This work was supported as a part of NCCR Automation, 
    a National Centre of Competence in Research, 
    funded by the Swiss National Science Foundation (grant number 51NF40\_225155). 
    \textsuperscript{1}Automatic Control Laboratory (IfA), 
    ETH Z\"urich, 8092 Z\"urich, Switzerland 
    \texttt{\small$\{$selfeki,rzuliani,nikschmid,lygeros$\}$@ethz.ch}. 
    \textsuperscript{3}Control and Automation Group, inspire AG, 
    8005 Z\"urich, Switzerland. 
    \texttt{\small efe.balta@inspire.ch}.
  }
}

\maketitle

\begin{abstract}
Applying policy optimization to Model Predictive Control (MPC) yields high-performance and reliable controllers. However, the resulting controllers often overfit their training conditions and suffer significant performance degradation in unseen tasks.
We propose a novel framework combining policy optimization with meta-learning to train highly adaptable MPC controllers.
Our approach enables rapid adaptation to unseen tasks, maintaining high performance at a fraction of the computational cost required for full retraining.
Furthermore, we integrate system identification into the pipeline to continuously refine both the MPC hyperparameters and the underlying predictive models.
We validate our proposed methodology on a Ball-on-Plate system, demonstrating superior adaptability across various parameterized trajectory-tracking tasks.
\end{abstract}


\section{Introduction}

Model Predictive Control (MPC) is a popular optimal control technique favored for its flexibility, ease of use, and constraint-handling capabilities.
Its performance relies primarily on an accurate system model for reliable open-loop predictions, as well as well-designed cost functions and constraints to meet performance and safety requirements.

Recently, gradient-based policy optimization has emerged as a principled approach to tuning MPCs for improved performance and guaranteed constraint satisfaction \cite{Amos2018,zuliani2025bp}. 
In this framework, cost and constraints are treated as trainable variables, and gradient descent is used to optimize closed-loop performance. 
System identification can additionally be integrated within the training pipeline to jointly optimize the predictive model alongside the rest of the MPC parameters \cite{Zuliani2026}.
These methods, however, remain computationally expensive, requiring multiple closed-loop iterations to converge.
This bottleneck becomes particularly limiting in hardware implementations, where each iteration carries irreducible costs in terms of execution time and system reset overhead.
Furthermore, the resulting controller is optimized for specific training conditions and may fail to generalize if these change.

Meta-learning is a class of algorithms that aim to learn policies that generalize across multiple tasks, enabling rapid adaptation to unseen tasks using limited data.
To overcome the computational bottleneck associated with traditional closed-loop tuning, we leverage meta-learning techniques to efficiently optimize MPC designs across a diverse range of problem definitions.
Specifically, we propose an end-to-end MPC design pipeline that incorporates meta-learning, online system identification and gradient-based parameter tuning by adapting the model-agnostic meta-learning (MAML) algorithm \cite{finn2017model} to the policy optimization framework introduced in \cite{Zuliani2026}.
To test the practical viability of our approach, we deploy it on a Ball-on-Plate (BoP) system, and demonstrate rapid adaptation to varying reference trajectories.
Our experiments provide strong evidence that 1) the tuning method of \cite{Zuliani2026} effectively improves the performance of an MPC controller on a hardware platform; 2) combining system identification with gradient-based tuning leads to a significant improvement in convergence speed at no additional computational cost; and 3) the proposed framework successfully leverages MAML to adapt to varying reference trajectories in a real-world setting with significantly reduced computational overhead.

The remainder of this paper is structured as follows: \cref{section:prob} describes the problem formulation, \cref{section:method} introduces our method, and \cref{section:exp} presents the hardware experiments.

\subsubsection*{Related work} To the best of our knowledge, applying policy optimization to MPC was first proposed in \cite{Amos2018}, and subsequently in \cite{Agrawal2019}. Since then, this method has been extended to uncertain systems \cite{Zuliani2025}, fully unknown systems \cite{Zuliani2026l4dc}, and has been combined with system identification to allow for concurrent optimization of the prediction model and the MPC parameters \cite{Zuliani2026}. 
In \cite{zuliani2025bp}, the authors propose a framework, rooted in nonsmooth analysis, to derive rigorous convergence guarantees.
A separate, yet highly relevant direction, proposes combining MPC controllers with reinforcement learning, using the optimal cost attained by the MPC as a value function approximator (see \cite{Gros2020} and subsequent works by the same authors).

Even though MAML is recognized as an effective approach for fast parameter adaptation, its use in MPC settings is generally limited to adapting the system model within the MPC framework.
For example, \cite{mei2025fast, yan2024mpc, Yan2026} use MAML to improve prediction accuracy of offline-trained neural network representations of the system model, mitigating the sim2real gap. In contrast, we shift the focus from the model to the controller by asking the following question: can we find MPC parameters that rapidly adapt to high performance on unseen tasks, by training the policy on the real system across a limited number of tasks?
To the best of our knowledge, no prior research explores the integration of MAML for task-specific adaptation of MPC controllers' objective and constraint definitions.

\subsubsection*{Notation} $\mathbb{Z}, \mathbb{N}, \mathbb{R}$ denote the sets of integer, natural, and real numbers, respectively. $k \in \mathbb{Z}_{[a,b)}$ defines an integer $k$, such that $k \in \{a, a + 1, \dots, b - 1 \}$ where $a < b$. $I_n$ is an identity matrix of dimension $n \times n$. $Q := L(p)L^\top(p)$ defines a symmetric positive-definite matrix $Q$ parameterized by parameter vector $p$ through the Cholesky decomposition of a lower triangular matrix $L(p)$, whereas $Q := \text{diag}(p)$ is a matrix $Q$ with parameter vector $p$ as diagonal entries. $Q \succ 0$ describes a positive-definite matrix, and $P_\mathcal{P}$ is the orthogonal projector to a convex set $\mathcal{P} \subset \mathbb{R}^{n}$. By default, $\| \cdot \|$ denotes the (induced) $2$-norm of a (matrix) vector. When a subscript $i \in \{1,2,F\}$ is used, $\| \cdot \|_i$ denotes the corresponding $i$-norm or the Frobenius norm, respectively.
\section{Problem Formulation}\label{section:prob}

Consider a dynamical system consisting of a plant $f$ and an MPC controller over a finite horizon $t\in\Z_{[0,T)}$
\begin{subequations}
\label{eq:prob:system}\begin{align}
x_{t+1} & = f(x_t, u_t), \label{eq:prob:system:f} \\
y_t & = \mpc(x_t, y_{t-1}, r_{\cdot|t}; p, \theta), \\
u_t & = \Pi_{u_0}(y_t)
\end{align}
\end{subequations}
where $x_t \in \R^{n_x}$, $u_t \in \R^{n_u}$, $r_{\cdot|t} \in \R^{n_r}$, and $y_t \in \R^{Tn_u}$ denote the state, input, MPC reference and MPC output at time $t$, respectively. 
The state is assumed to be fully measurable.  
Note that $u_t$ is contained in $y_t$, and $\Pi_{u_0}(\cdot)$ projects the MPC solution onto its first planned input.
In addition to tracking the reference, we assume that desired state and input constraints
\begin{align} \label{eq:prob:constraints}
\begin{split}
H_x x_t &\leq h_x,~ \forall t\in\Z_{[0,T]}, \\ H_u u_t &\leq h_u,~ \forall t\in\Z_{[0,T)}.
\end{split}
\end{align}
are given as part of the MPC design specification.
We assume that $f$ is unknown but a model of the form
\begin{align}\label{eq:prob:plant}
f_\theta(x,u) = \varphi(x,u) + \theta^{\top} \phi(x,u),
\end{align}
is available, where $\varphi:\R^{n_x} \times \R^{n_u}\to \R^{n_x}$ and $\phi:\R^{n_x} \times \R^{n_u} \to \R^{n_\theta}$ are functions and $\theta\in\R^{n_\theta \times n_x}$ is a trainable parameter. 
At each time-step, the MPC output $y_t$ is obtained by solving the following optimization problem, designed to promote tracking of the reference $r$ while satisfying the constraints in \cref{eq:prob:constraints}:
\begin{subequations}
\label{eq:prob:mpc}\begin{align}
\operatorname*{minimize}_{y_t=(x_{\cdot|t},u_{\cdot|t},\epsilon_{\cdot|t})} & \quad %
\ell_\text{slack}(\epsilon_{\cdot|t}) + \ell_\text{track}(x_{\cdot|t},u_{\cdot|t},r_{\cdot|t};p) \label{eq:MPC_cost} \\
\text{subject to}~~ %
& \quad x_{i+1|t} = A_{i}(y_{t-1},\theta) x_{i|t} + B_{i}(y_{t-1},\theta) u_{i|t} \notag \\ 
& \hphantom{~~ x_{i+1|t} =} \hspace{0.5em} + c_{i}(y_{t-1},\theta), \ i \in \Z_{[0,N)}, \label{eq:prob:mpc:dynamics} \\
& \quad H_x x_{i|t} \le h_x + \epsilon_{i|t}, \ i \in \Z_{[0, N]}, \label{eq:prob:mpc:state_constraints} \\
& \quad H_u u_{i|t} \le h_u, \ i \in \Z_{[0,N)}, \label{eq:prob:mpc:input_constraints} \\
& \quad \epsilon_{i|t} \geq 0, \ i \in \Z_{[0, N]}, \\
& \quad x_{0|t} = x_t. \label{eq:prob:mpc:initial_state}
\end{align}
\end{subequations}
The decision variable in \cref{eq:prob:mpc} is a prediction of the future state and input trajectories $x_{\cdot|t}:=(x_{0|t},\dots,x_{N|t})$, $u_{\cdot|t}:=(u_{0|t},\dots,u_{N-1|t})$ and a sequence of non-negative slack variables $\epsilon_{\cdot|t}:=(\epsilon_{0|t},\dots,\epsilon_{N|t})$ used for relaxing the state constraints in \cref{eq:prob:mpc:state_constraints} to ensure feasibility.
The dynamics are obtained by linearizing the model $f_\theta$ along the previous MPC solution $(x_{\cdot|t-1},u_{\cdot|t-1})$, shifted by one time-step
\begin{subequations}\label{eq:prob:linearized_dynamics}
\begin{align}
A_{i}(y_{t-1},\theta) & = \nabla_x f_\theta(x_{i+1|t-1},u_{i+1|t-1}),\\
B_{i}(y_{t-1},\theta) & = \nabla_u f_\theta(x_{i+1|t-1},u_{i+1|t-1}),\\
c_{i}(y_{t-1},\theta) & = f_\theta(x_{i+1|t-1}, u_{i+1|t-1}) - A_{i}(y_{t-1},\theta) x_{i+1|t-1} \notag \\ & ~~~ - B_{i}(y_{t-1},\theta) u_{i+1|t-1},
\end{align}
\end{subequations}
where $u_{N|t-1}=u_{N-1|t-1}$. By employing this linearization strategy, which resembles the real-time iteration of \cite{Gros2016}, we can obtain accurate prediction models without having to resort to nonlinear optimization \cite{zuliani2025bp}.
The slack penalty cost is
\begin{align*}
\ell_\text{slack}(\epsilon_{\cdot|t}) & := c_1 \| \epsilon_{\cdot|t} \|_2^2 + c_2 \| \epsilon_{\cdot|t} \|_1,
\end{align*}
for some $c_1,c_2>0$, and the tracking cost is
\begin{align*}
\ell_\text{track}(x_{\cdot|t},u_{\cdot|t},r_{\cdot|t};p) & := \|x_{N|t}-r_{N|t}\|_{P_x(p)}^2 \notag \\ + & \sum_{i=0}^{N-1} \|x_{i|t}-r_{i|t}\|_{Q_x(p)}^2 + \|u_{i|t}\|_{R_u(p)}^2,
\end{align*}
where $r_{\cdot|t} = (r_{0|t},\dots,r_{N|t})$, with $r_{t+i}=r_{i|t}$ for all $t\in\Z_{[0,T)}$ and $i\in\Z_{[0,N]}$. 
The cost matrices $P_x(p)$, $Q_x(p)$, and $R_u(p)$ depend on a trainable parameter $p\in\R^{n_p}$.

Our goal is to solve the following nonsmooth and nonconvex policy optimization problem for multiple values of $r$
\begin{subequations}\label{eq:prob:policy_optimization}
\begin{align}
\operatorname*{minimize}_{p,x,u,y} & ~~ \mathcal{C}_\text{track}(x, r) \\
\text{subject to} %
& ~~ x_{t+1} =  f(x_t, u_t), ~ t \in \Z_{[0,T)}, \\
& ~~ y_t = \mpc (x_t,y_{t-1},r_{\cdot|t};p,{\theta_r}), ~ t \in \Z_{[0,T)}, \\
& ~~ u_t = \Pi_{u_0}(y_t), ~ t \in \Z_{[0,T)}, \\
& ~~ H_x x_t \leq h_x, ~ t \in \mathbb{Z}_{[0,T]}, \\
& ~~ H_u u_t\leq h_u, ~ t \in \mathbb{Z}_{[0,T)}, \label{eq:state-input_constraints} \\
& ~~ x_0, y_{-1}, \text{ and } \theta_r \text{ given},
\end{align}
\end{subequations}
where $\mathcal{C}_\text{track}(x,r)$ is the tracking cost. 
At the same time, we wish to choose a value of ${\theta_r}$ so that $f_{\theta_r}(x_t,u_t) \approx f(x_t,u_t)$ for all $t$ by using data collected from the real system.
We assume that the system is repeatedly executed from the same initial state $x_0$ for iterations of $T$ time-steps.
Since we do not assume that there exists a $\theta^*$ for which $f_{\theta^*}=f$, we consider different parameters $\theta_r$ for each $r$, to maximize the accuracy of the model for each task.

We consider the problem of solving \cref{eq:prob:policy_optimization} for several values of $r$, a situation in which the method in \cite{Zuliani2026} becomes computationally intractable.
The next section outlines our proposed solution.
\section{Methodology}\label{section:method}

To reduce the computational effort required to obtain an optimal design $p_r$ for an unseen new reference $r$, we adapt the \emph{MAML algorithm} \cite{finn2017model} to the MPC policy optimization framework of \cite{zuliani2025bp,Zuliani2026}. 
Our goal becomes obtaining a parameter $\bar{p}$ from which any $p_r$ can be retrieved with only a few updates, keeping the interaction with the system at a minimum. 
We first rewrite \cref{eq:prob:policy_optimization} as an unconstrained problem, amenable to gradient descent. 
We consider the following reformulation
\begin{align}\label{eq:method:policy_optimization_unconstrained}
\operatorname*{min.}_{p} ~ \mathcal{C}_\text{track}(x(r,p,\theta_r),r) \! + \! \mathcal{C}_{\text{slack}}(\epsilon(r,p,\theta_r)) =: \mathcal{C}_\text{full}(r,p,\theta_r),
\end{align}
where $x(r,p,\theta_r):=\{x_t(r,p,\theta_r)\}_{t\in\Z_{[0,T]}}$ and $\epsilon(r,p,\theta_r):=\{\epsilon_t(r,p,\theta_r)\}_{t\in\Z_{[0,T)}}$ denote the closed-loop state and slack trajectories satisfying \cref{eq:prob:system} and \cref{eq:prob:mpc} for $t\in\Z_{[0,T]}$, and $\mathcal{C}_{\text{slack}}(\epsilon):= c_3 \| \epsilon \|$, with $c_3>0$.
Note that \cref{eq:state-input_constraints} is always satisfied thanks to the MPC constraints \cref{eq:prob:mpc:input_constraints}, and can therefore be omitted from \cref{eq:method:policy_optimization_unconstrained}. 
Evaluating $x(r,p,\theta_r)$, $u(r,p,\theta_r)$ and $\epsilon(r,p,\theta_r)$ requires access to the real system, whereas the MPC is utilizing the model \cref{eq:prob:plant} with nominal parameter $\theta_r$.

The MAML algorithm can then be applied to obtain a parameter $\bar{p}$ for which the solution of \cref{eq:method:policy_optimization_unconstrained} can be obtained for any $r$ by running a single gradient descent step on \cref{eq:method:policy_optimization_unconstrained}, that is
\begin{align}\label{eq:method:task_specific_parameter}
p_r \approx p_r(\bar{p}) := \bar{p} - \alpha \nabla_p \mathcal{C}_{\text{full}}(r,\bar{p},\theta_r),
\end{align}
where $\alpha>0$ is a positive stepsize. 
We call $p_r(\bar{p})$ the \emph{task-specific parameter}, and $\bar{p}$ the \emph{generalized parameter}.
To obtain $\bar{p}$ we solve the following problem
\begin{align}\label{eq:method:maml_objective}
\operatorname*{minimize}_{\bar{p}} ~ \sum_{r\in \mathcal{R}} \mathcal{C}_{\text{full}}(r,p_r(\bar{p}),\theta_r) =: \bar{\mathcal{C}}(\bar{p}),
\end{align}
where $\mathcal{R}$ is a finite set of references. 
The goal of \cref{eq:method:maml_objective} is to find a generalized parameter $\bar{p}$ such that each adapted $p_r(\bar{p})$ effectively minimizes its respective cost $\mathcal{C}_{\text{full}}(r,\cdot,\theta_r)$.
Consequently, after training, near optimal parameters for unseen references can be obtained with a single closed-loop interaction via \cref{eq:method:task_specific_parameter}.
To obtain the gradient of \cref{eq:method:maml_objective}, we apply the chain rule leveraging the definition of $p_r(\bar{p})$ in \cref{eq:method:task_specific_parameter}:
\begin{align}\label{eq:method:maml_objective_gradient:exact}
\nabla \bar{\mathcal{C}}(\bar{p}) & \stackrel{\phantom{(8)}}{=} \sum_{r\in \mathcal{R}} \nabla_{\bar{p}} \mathcal{C}_\text{full}(r,p_r(\bar{p}),\theta_r) \notag \\
& \stackrel{\phantom{(8)}}{=} \sum_{r\in \mathcal{R}} \nabla p_r(\bar{p})^\top \nabla_{p} \mathcal{C}_\text{full}(r,p_r(\bar{p}),\theta_r) \notag \\
& \stackrel{\cref{eq:method:task_specific_parameter}}{=} \sum_{r\in\mathcal{R}} [I - \alpha \nabla_p^2 \mathcal{C}_{\text{full}}(r,\bar{p},\theta_r)]^{\top} \nabla_{p} \mathcal{C}_{\text{full}}(r,p_r(\bar{p}),\theta_r).
\end{align}
Using \cref{eq:method:maml_objective_gradient:exact} directly to solve \cref{eq:method:maml_objective} is challenging for two reasons: 1) obtaining an exact Jacobian is computationally expensive as it requires the computation of second derivatives of the MPC function, and 2) the true system is unknown, hence so is the true gradient.
Instead, we propose combining system identification and an approximation of \cref{eq:method:maml_objective_gradient:exact} that does not require second derivatives.
The method is summarized in \cref{alg:method:maml}, where $k$ denotes the iteration, $x_r^k$ and $u_r^k$ denote $x(r,p_r^k,\theta_r^k)$ and $u(r,p_r^k,\theta_r^k)$, respectively, with $p_r^k:=p_r(\bar{p}^k)$, and $\alpha_k$ is a vanishing stepsize, such as $\alpha_k=c\frac{\log{(k+2)}}{k+1}$ for some $c>0$. 
Estimating $\nabla \bar{\mathcal{C}}(\bar{p})$ requires two system interactions: the first to form $\nabla_p \mathcal{C}_\text{full}(r,\bar{p},\theta_r)$ and obtain $p_r(\bar{p})$, the second to evaluate $\nabla_{p} \mathcal{C}_{\text{full}}(r,p_r(\bar{p}),\theta_r)$.

\begin{algorithm}[ht!]
    \crefname{algorithm}{Alg.}{Algs.}
    \caption{MAML for Differentiable MPC} \label{alg:method:maml}
    \begin{algorithmic}[1]
        \REQUIRE $\{\alpha_k\}_{k \in \mathbb{N}} \subset \R_{>0}, \mathcal{R}$, $\theta^{0}$, $\bar{p}^0$.
        \STATE \textbf{Init}: $\theta_r^0\!\gets\!\theta^0$, $A^0_r \!\gets \!\lambda I$, $b_r^0 \!\gets \!\lambda \theta^0_r$, $H_r^0 \!\gets \!I$ for all $r\in\mathcal{R}$.
        \STATE \textbf{Init}: $k \gets 0$.
        \WHILE{not converged}
            \FORALL {$r \in \mathcal{R}$}
                \STATE Obtain $\hat{\nabla}_p \mathcal{C}_\text{full}(r,\bar{p}^k,\theta_r^k)$ via \cref{alg:method:bpmpc}.
                \STATE Form $p_r^k=p_r(\bar{p}^k)$ via \cref{eq:method:task_specific_parameter}.
                \STATE Obtain $x_r^k := x(r,p^k_r,\theta_r^k)$ and $u_r^k := u(r,p^k_r,\theta_r^k)$.
                \STATE Obtain $\theta_r^{k+1}$, $A^{k+1}_r$, and $b^{k+1}_r$ via \cref{alg:method:rls}.
                \STATE Estimate $\hat{\nabla}_p \mathcal{C}_{\text{full},r}^k\approx\nabla_p \mathcal{C}_\text{full}(r,p^k_r,\theta^k_r)$ via \cref{alg:method:bpmpc}.
            \ENDFOR
        \STATE Estimate $\hat{\nabla} \bar{\mathcal{C}}^k \approx \nabla \bar{\mathcal{C}}(\bar{p}^k)$ via \cref{eq:method:maml_objective_gradient:inexact_bfgs}.
        \STATE Update $\bar{p}^{k + 1}$ via $\bar{p}^{k+1} = \bar{p}^k - \alpha_k \hat{\nabla} \bar{\mathcal{C}}^k$.
        \STATE $k \gets k + 1$.
        \ENDWHILE
        \RETURN $\bar{p}^k$.
    \end{algorithmic}
\end{algorithm}

Next, we outline how system identification can be used to update $\theta_r^k$ (\cref{section:method:sys_id}), how to estimate $\nabla_p \mathcal{C}_\text{full}(r,p,\theta)$ given any $p$ and $\theta$ with imperfect model knowledge (\cref{section:method:bpmpc}), and how to approximate $\nabla \bar{\mathcal{C}}(\bar{p}^k)$ (\cref{section:method:maml}).

\subsection{System Identification}\label{section:method:sys_id}


For a given task $r$, we can leverage knowledge of $x_r^k$ and $u_r^k$ for all $k$ to update the system estimate $\theta_r^k$ to ensure that $f(x_{t,r}^k,u_{t,r}^k) \approx f_{\theta_r^k}(x_{t,r}^k,u_{t,r}^k)$ for all $t\in\Z_{[0,T)}$ after sufficiently many episodes, where $x_{t,r}:=x_t(r,p_r^k,\theta_r^k)$ and $u_{t,r}:=u_t(r,p_r^k,\theta_r^k)$, with $p_r^k:=p_r(\bar{p}^k)$. 
Because \cref{eq:prob:plant} is parameter affine, we can rewrite \cref{eq:prob:system:f} as
\begin{align}\label{eq:method:sys_id_equation}
z_{t,r}^k = \theta_{r}^{k,\top} \psi_{t,r}^k,
\end{align}
where $\psi_{t,r}^k = \phi(x_{t,r}^k,u_{t,r}^k)$ and $z_{t,r}^k=x_{t+1,r}^k-\varphi(x_{t,r}^k,u_{t,r}^k)$.
To identify $\theta_{r}^k$ we employ a recursive least-squares estimator \cite{abbasi2011improved}
\begin{subequations}
\label{eq:method:rls}\begin{align}
A^{k+1}_r & = A^k_r + {\textstyle\sum}_{t=0}^{T-1} \psi_{t,r}^k \psi_{t,r}^{k,\top}, \label{eq:method:rls:A} \\
b^{k+1}_r & = b^k_r + {\textstyle\sum}_{t=0}^{T-1} \psi_{t,r}^k z_{t,r}^{k,\top}, \label{eq:method:rls:b} \\
\theta^{k+1}_r & = (A^{k+1}_r)^{-1} b^{k+1}_r, \label{eq:method:rls:theta}
\end{align}
\end{subequations}
with $A^0_r=\lambda I$, $b^0_r=\lambda \theta_r^0$ for some initial estimate $\theta^0_r$ and regularization parameter $\lambda>0$.
Observe that the update of $\theta_r^k$ only occurs at the end of each iteration using the data collected from the real system.
The recursive system identification procedure is summarized in \cref{alg:method:rls}.
As we show in \cref{section:method:bpmpc}, system identification not only improves the prediction model of the MPC (used for the linearized dynamics in \cref{eq:prob:linearized_dynamics}) but also the estimate of the gradient $\nabla_p \mathcal{C}_\text{full}(r,p_r^k,\theta_r^k)$.

\begin{algorithm}[ht!]
    \caption{Recursive Least Squares Update}
    \label{alg:method:rls}
    \begin{algorithmic}[1]
        \REQUIRE Trajectories $x_r^k$, $u_r^k$, current matrices $A^k_r$ and $b^k_r$.
        \STATE Compute $z_{t,r}^k = x_{t+1,r}^k - \varphi(x_{t,r}^k,u_{t,r}^k)$ for all $t\in\mathbb{Z}_{[0,T)}$.
        \STATE Compute $\psi_{t,r}^k = \phi(x_{t,r}^k,u_{t,r}^k)$ for all $t\in\mathbb{Z}_{[0,T)}$.
        \STATE Compute $A^{k+1}_r$ and $b^{k+1}_r$ using \cref{eq:method:rls:A,eq:method:rls:b}.
        \STATE Compute $\theta^{k+1}_r$ using \cref{eq:method:rls:theta}.
        \RETURN $\theta^{k+1}_r$, $A^{k+1}_r$, $b^{k+1}_r$.
    \end{algorithmic}
\end{algorithm}

\subsection{Closed-Loop Gradient Estimation}\label{section:method:bpmpc}

For each iteration $k$ and reference $r$, \cref{alg:method:maml} requires estimating two closed-loop gradients $\nabla_p \mathcal{C}_\text{full}(r,\bar{p}^k,\theta_r^k)$ and $\nabla_p \mathcal{C}_\text{full}(r,p_r^k,\theta_r^k)$. 
To obtain an estimate $\hat{\nabla} \mathcal{C}_{\text{full}}(r,p,\theta)$ of $\nabla_p \mathcal{C}_\text{full}(r,p,\theta)$ for any generic $r$, $p$, and $\theta$, we use the \emph{Backpropagation-MPC} (\emph{BP-MPC}) automatic differentiation methodology proposed in \cite{zuliani2025bp}. 
BP-MPC is a framework tailored for sensitivity computation in policy optimization problems involving MPC controllers.
At its core, BP-MPC relies on the application of the chain rule to the cost function
\begin{multline}\label{eq:method:bpmpc_chain_rule_cost}
\hat{\nabla} \mathcal{C}_{\text{full}}(r,p,\theta) = \nabla_x \mathcal{C}_\text{track}(x(r,p,\theta),r) \hat{\nabla}_p x(r,p,\theta) \\ + \nabla \mathcal{C}_\text{slack}(\epsilon(r,p,\theta)) \hat{\nabla}_p \epsilon(r,p,\theta).
\end{multline}
The gradients $\nabla_x \mathcal{C}_\text{track}(x,r)$ and $\nabla \mathcal{C}_\text{slack}(\epsilon)$ can be obtained from elementary algebraic calculations using the definitions of $\mathcal{C}_\text{track}$ and $\mathcal{C}_\text{slack}$ thanks to the knowledge of $x(r,p,\theta)$ and $\epsilon(r,p,\theta)$. 
On the other hand $\nabla_p x(r,p,\theta)$ and $\nabla_p \epsilon(r,p,\theta)$ are unknown due to the unknown $f$. Given a model $f_\theta$ of $f$, we estimate $\nabla_p x(r,p,\theta)$ and $\nabla_p \epsilon(r,p,\theta)$ as follows
\begin{subequations}
\label{eq:method:bpmpc_recursion_inexact}\begin{flalign}
\hat{\nabla}_p x_{t+1}(r,p,\theta) & = \nabla_x f(x_t,u_t,\theta) \hat{\nabla }_px_{t}(r,p,\theta) \notag \\  & \hspace{-0.6cm} + \nabla_u f(x_t,u_t,\theta) \hat{\nabla}_p u_{t}(r,p,\theta), && \\
\hat{\nabla}_p y_{t}(r,p,\theta) & = \nabla_{x}\mpc(x_t,y_{t-1},r_t;p,\theta) \hat{\nabla}_p x_{t}(r,p,\theta) \notag \\ & \hspace{-0.6cm} + \nabla_y \mpc (x_t,y_{t-1},r_t;p,\theta) \hat{\nabla}_p y_{t-1}(r,p,\theta)  \notag \\ & \hspace{-0.6cm} + \nabla_p \mpc(x_t,y_{t-1},r_t;p,\theta), \\
\hat{\nabla}_p u_{t}(r,p,\theta) & = \Pi_{u_0} (\hat{\nabla}_p y_{t}(r,p,\theta)),
\end{flalign}
\end{subequations}
where $\hat{\nabla}_p x_{0}(r,p,\theta)=\hat{\nabla}_p y_{-1}(r,p,\theta)=0$ since the initial condition and linearization trajectory do not depend on $p$.

To obtain the derivative of the MPC map, BP-MPC applies a nonsmooth implicit function theorem to a dual-based optimality condition of the problem. 
We refer the reader to \cite{zuliani2025bp} for further details.
Under the mild conditions outlined in \cite{zuliani2025bp}, gradients are guaranteed to exist almost everywhere, and wherever they do not exist, BP-MPC provides a subgradient which can be used instead without hindering the policy optimization procedure.
The entire gradient estimation procedure is summarized in \cref{alg:method:bpmpc}.
The algorithm is guaranteed to converge to a critical point provided that the closed-loop trajectories are sufficiently rich for system identification, and there exists some $\theta^*$ such that $f=f_{\theta^*}$. 
Because we do not impose this latter assumption, a detailed analysis of this convergence falls outside the scope of this paper. 
For a comprehensive discussion, we refer the reader to \cite{Zuliani2026}.

\begin{algorithm}
    \caption{Sensitivity computation with BP-MPC}
    \label{alg:method:bpmpc}
    \begin{algorithmic}[1]
        \REQUIRE $r$, $p$, $\theta$.
        \STATE \textbf{Init}: $\hat{\nabla}_p x_{0} = \hat{\nabla}_p y_{-1} = 0$, $x_0$.
        \FOR{$t \in \mathbb{Z}_{[0, T)}$}
            \STATE Fetch $y_{t}$ and $u_t=\Pi_{u_0}(y_t)$ from the solution of \eqref{eq:prob:mpc}.
            \STATE Fetch $x_{t+1}$ from the recorded trajectory $x(r, p, \theta)$.
            \STATE Compute $\hat{\nabla}_p x_{t+1}$, $\hat{\nabla}_py_{t}$, and $\hat{\nabla}_p u_{t}$ using \cref{eq:method:bpmpc_recursion_inexact}.
        \ENDFOR
        \STATE Compute $\hat{\nabla} \mathcal{C}_{\text{full}}(r,p,\theta)$ using \cref{eq:method:bpmpc_chain_rule_cost}.
        \RETURN $\hat{\nabla}_p\mathcal{C}_{\text{full}}(r,p,\theta)$.
    \end{algorithmic}
\end{algorithm}

\subsection{Model Agnostic Meta-Learning}\label{section:method:maml}

Given an estimate of $\nabla_p \mathcal{C}_\text{full}(r,\bar{p}^k,\theta^k_r)$ for each $r$, we now focus on approximating the gradient of the objective in \cref{eq:method:maml_objective}, avoiding the exact expression in \cref{eq:method:maml_objective_gradient:exact} which requires computationally expensive second derivatives.
One way to do this is to use a Quasi-Newton BFGS approximation of the Hessian \cite{wright1999numerical}, defined as

\begin{subequations}
\label{eq:method:maml_objective_gradient:inexact_bfgs}\begin{align}
\nabla \bar{\mathcal{C}}(\bar{p}^k) \approx \sum_{r\in\mathcal{R}} (I - \alpha H_r^k)^\top \nabla_{p} \mathcal{C}_{\text{full}}(r,p_r^k,\theta_r^k),
\end{align}
where $H_r^k$ approximates the true Hessian $\nabla^2_p \mathcal{C}_\text{full}(r,\bar{p}^k,\theta_r^k)$ and is defined iteratively via the direct BFGS update:
\begin{align}
H^{k+1}_r & = H^k_r + \frac{y_r^k y_r^{k,\top}}{y_r^{k,\top} s_r^k} - \frac{H_r^k s_r^k s_r^{k,\top} H_r^k}{s_r^{k,\top} H_r^k s_r^k}, \\
s_r^k & = p^{k+1}_r - p^k_r, \\
y_r^k & = \nabla_p \mathcal{C}_{\text{full}}(r,p_r^{k+1},\theta_r^{k+1}) - \nabla_p \mathcal{C}_{\text{full}}(r,p_r^k,\theta_r^k).
\end{align}
\end{subequations}
In our framework, using BFGS over standard gradient descent introduces no additional computational overhead, as evaluating the gradient $\nabla_{p}\mathcal{C}_\text{full}$ remains the primary computational bottleneck.
\section{Experiments}\label{section:exp}

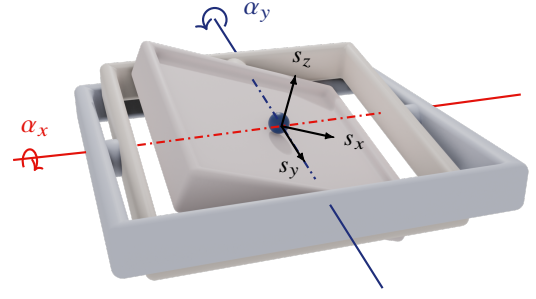
\begin{figure}
\centering
\begin{tikzpicture}
    \node[anchor=south west, inner sep=0pt] (image) at (0,0) {
        \includegraphics{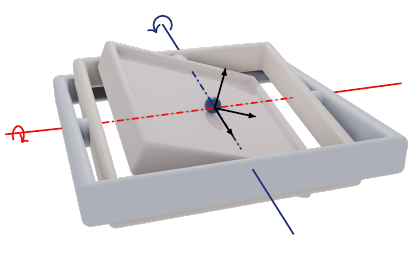}
    };
    \begin{scope}[x={(image.south east)}, y={(image.north west)}]
        \node[text=red1, anchor=east] at (0.1, 0.575) {$\alpha_x$};
        \node[text=blue2, anchor=west] at (0.425, 0.925) {$\alpha_y$};
        \node[anchor=west] at (0.515, 0.782) {\small$s_z$};
        \node[anchor=center] at (0.65, 0.535) {\small$s_x$};
        \node[anchor=west] at (0.49, 0.45) {\small$s_y$};
    \end{scope}
\end{tikzpicture}
\caption{Simplified depiction of the Ball-on-Plate system.}
\label{schematic_bop}
\end{figure}

\begin{figure}
\centering
\includegraphics[width=0.8\columnwidth]{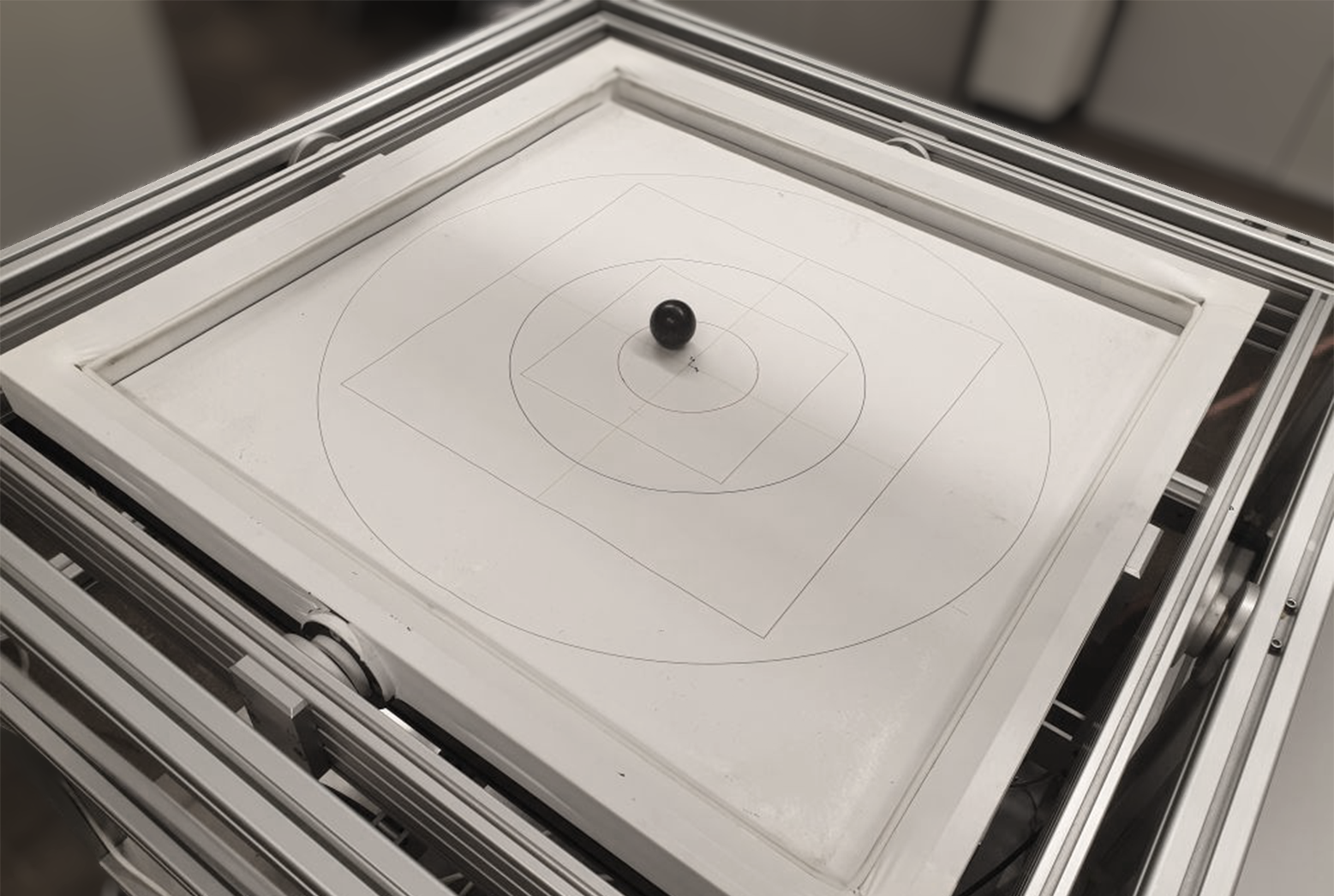}
\caption{The IfA hardware Ball-on-Plate control testbed; a motorized, ball-balancing plate.}
\label{fig_bop}
\end{figure}

We experimentally study the performance of our approach on the Ball-on-Plate testbed, shown in \cref{schematic_bop,fig_bop}. The system consists of a ball on a motorized screen. The ball-position is measured using a camera. A microcontroller transmits this data over a wireless network and accepts setpoints for the plate angles which it realizes via the motors. Our MPC is executed on a desktop computer with an i7-1165G7 CPU. 

We denote the state of the ball comprising its position and velocity by $x=[s_x~v_x~s_y~v_y]^\top$ and the input comprising tilt angles by $u= [\alpha_x~\alpha_y]^\top$. The tilt angle of the plate generates forces $g/\left(m+Ir^{-2}\right)\sin(\alpha_x)$ in the $x$ and respectively in the $y$ direction, where $g=\SI{9.81}{\meter\per{\second\squared}}$, $r$ is the ball radius, $m$ its mass, and $I$ its moment of inertia. The ball is further affected by friction. Neglecting slipping and viscous effects, we approximate the static and Coulomb friction by a force opposing the ball velocity with amplitude $F_c + (F_s-F_c)e^{-\frac{v}{v_s}}$, where $v=\sqrt{v_x^2+v_y^2}$, $F_i=\mu_img\cos{\alpha_x}\cos{\alpha_y}$ for $i\in\{s,c\}$ and $v_s$ and $\mu_i$ are friction coefficients \cite{olsson1998friction}. At every time-step $i$, we approximate the nonlinear dynamics by a linear time-varying model for the time-steps $i,\dots,i+N-1$ by linearizing the angle-induced forces around $\alpha_x,\alpha_y=0$, and friction-induced forces around the MPC solution at the previous time-step, $u_{i-1|i-1},\dots,u_{i+N-1|i-1}$ (or zero-angles, and almost-zero velocities at $i=0$). This leads to the Euler-discretized linear time-varying dynamics 
\begin{align}
        x_{t + 1} &= \begin{bmatrix} 1 & \Delta &  0 & 0 \\ 0 & 1 & 0 & 0  \\ 
        0 & 0 & 1 & \Delta \\ 0 & 0 & 0 & 1 \end{bmatrix} x_t + \Delta\begin{bmatrix}
            0 \\ \theta_I \\ 0 \\ \theta_I
        \end{bmatrix}u_t + \begin{bmatrix}
            0 \\ c_x \\ 0 \\ c_y
        \end{bmatrix},\quad\quad\quad~
    \label{eq:linear_bop_system}
\end{align}
where $\Delta=\SI{0.025}{\second}$ is the sampling-time, and
\begin{align*}
    c_x &= \theta_s \cos{\alpha_{x,t}}\cos{\alpha_{y,t}} + \theta_c v_{x,t} \cos{\alpha_{x,t}}\cos{\alpha_{y,t}}, \\
    c_y &=  \theta_s \cos{\alpha_{x,t}}\cos{\alpha_{y,t}} + \theta_c v_{y,t} \cos{\alpha_{x,t}}\cos{\alpha_{y,t}}, \\
    \theta_I & = \frac{mg}{\left( m + I/r^2 \right)}, \ \ \theta_s := \mu_s mg, \ \ \theta_c :=  \frac{mg}{v_s}(\mu_c - \mu_s).
\end{align*} 
The latter define the unknown physical parameters $\theta_I$, $\theta_s$, and $\theta_c$, which populate the non-zero entries of the parameter matrix $\theta$ identified via \cref{alg:method:rls}.

System \eqref{eq:linear_bop_system} is used as prediction model for the MPC with a prediction horizon of $N=20$ time-steps. We further impose 
\begin{equation*}
 \lvert s_x \rvert, \lvert s_y \rvert \le 0.3\mathrm{m}, \text{ and } \lvert \alpha_x \rvert, \lvert \alpha_y \rvert \le 11^\circ,
\end{equation*}
and parametrize the MPC cost matrices as $p := \{p_{P_x}, p_{Q_x}, p_{R_u}\}$, with $P_x := L(p_{P_x})L^\top(p_{P_x})$, $Q_x := \text{diag}(p_{Q_x})$, and $R_u := p_{R_u} I_2$. Unless stated otherwise, we initialize $p$ so that $Q_x = \text{diag}(10,1,10,1)$, $R_u:= 50 \cdot I_{2}$, and $P_x$ is the solution of the respective DARE. We choose
\begin{equation} \label{eq:BP-MPC_objective}
    \mathcal{C}_{\text{track}}(x, r) := \sum_{t=\tau}^{T} \lVert x_t - r_t \rVert^2_{Q},~~ \mathcal{C}_{\text{slack}}(\epsilon):= 0 \cdot \lVert \epsilon \rVert_1,
\end{equation}
where $Q := \text{diag}(100,0.01,100,0.01)$, $T=1000$, and $\tau=200$.
The tracking cost deliberately omits the first $\tau$ time-steps to prevent the large transient errors of navigating from the corner to the circular trajectory from skewing the total cost.

\begin{figure*}
    \centering
    \includegraphics{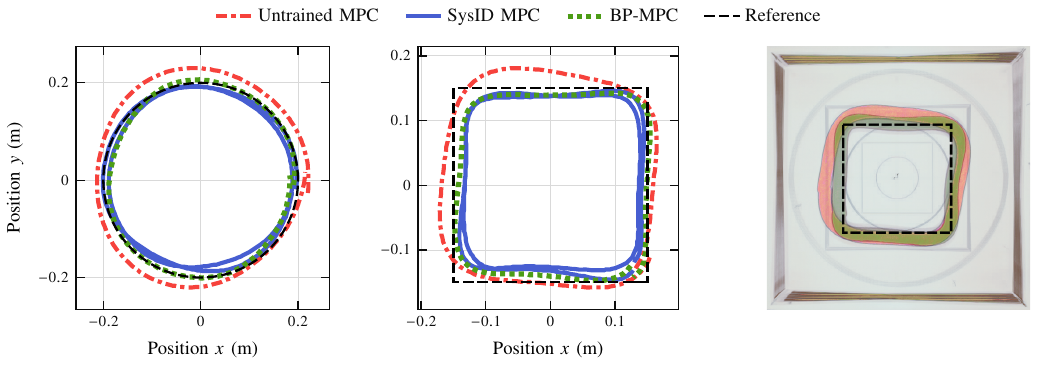}
    \caption{Closed-loop trajectories for the circle and square references generated by the initially untrained MPC, MPC with identified system parameters, and our BP-MPC approach. System identification and optimization of the MPC parameters both reduce tracking errors, see also Table \ref{tab:tracking_errors}. The right picture shows recorded ball trajectories for BP-MPC initially and after convergence; the dashed line is the square reference.}
    \label{fig:traj_plots}
\end{figure*}

\begin{figure}
    \centering
    \includegraphics{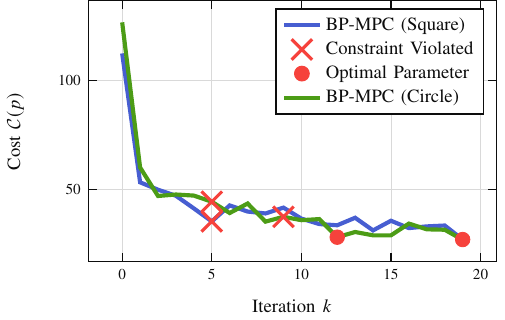}
    \caption{Closed-loop cost of the circle and square trajectory over episodes. Red Crosses: Constraint violations during episode; Red Balls: Episode with the least cost.}
    \label{fig_cost_over_iterations}
\end{figure}

We provide a numerical answer to three key questions, each with an individual experiment:
\begin{enumerate}[label=\Alph*.]
    \item Does the BP-MPC scheme improve closed loop performance over episodes?
    \item Does system identification improve performance of BP-MPC?
    \item Does warm-starting BP-MPC with MAML improve performance on unseen references?
\end{enumerate}
All experiments are implemented using CasADi with the DAQP solver \cite{Arnstrom2022}. Before the beginning of each episode, the plate is tilted towards the bottom left corner, ensuring that the ball always starts at the same initial position. We use two reference trajectories, (\textit{i}) a circle of radius $\SI{0.2}{\meter}$, and
(\textit{ii}) a square of side length $\SI{0.3}{\meter}$. Both references are repeated five times with $\SI{5}{\second}$ per cycle. 

\subsection{Does BP-MPC Improve MPC?}
We first study the impact of the BP-MPC tuning mechanism combined with system identification.
For this we run a simplified version of \cref{alg:method:maml} that does not include any meta learning, reported in \cref{alg:exp:maml_no_maml}.

\begin{algorithm}[ht!]
    \crefname{algorithm}{Alg.}{Algs.}
    \caption{Task-Specific BP-MPC} \label{alg:exp:maml_no_maml}
    \begin{algorithmic}[1]
        \REQUIRE $\{\alpha_k\}_{k \in \mathbb{N}} \subset \R_{>0}$, $r$, $\theta_r^{0}$, $p_r^0$.
        \STATE \textbf{Init}: $k \gets 0$, $A^0_r \gets \lambda I$, $b_r^0 \gets \lambda \theta^0_r$.
        \WHILE{not converged}
            \STATE Obtain $x_r^k := x(r,p^k_r,\theta_r^k)$ and $u_r^k := u(r,p^k_r,\theta_r^k)$.
            \STATE Obtain $\theta_r^{k+1}$, $A^{k+1}_r$, and $b^{k+1}_r$ via \cref{alg:method:rls}.
            \STATE Estimate $\hat{\nabla}_p \mathcal{C}_{\text{full},r}^k\approx\nabla_p \mathcal{C}_\text{full}(r,p^k_r,\theta^k_r)$ via \cref{alg:method:bpmpc}.
            \STATE Update $p_r^{k+1}$ via $p_r^{k+1} = p_r^k - \alpha_k \hat{\nabla}_p \mathcal{C}_{\text{full},r}^k$.
            \STATE $k \gets k + 1$.
        \ENDWHILE
        \RETURN $p_r^k$.
    \end{algorithmic}
\end{algorithm}

The tracking performance of the untrained MPC, untrained MPC with pre-identified $\theta$, and our BP-MPC scheme is depicted in Fig.~\ref{fig:traj_plots}, the decrease of closed-loop costs over iterations is depicted in Fig.~\ref{fig_cost_over_iterations}. The cost function, equivalent to the tracking cost, is listed in Tab.~\ref{tab:tracking_errors} with PID and LQR as additional benchmarks. 

\begin{table}[!tb]
    \centering
    \begin{tabular}{c c c c c l}
        \toprule
        & PID & LQR* & MPC & MPC* & BP-MPC \\
        \midrule
        \textbf{Circle} & 0.124 & 0.105 & 0.038 & 0.024 & \textbf{0.019} ($\SI{15}{\percent}$ PID costs) \\
        \textbf{Square} & 0.116 & 0.102 &  0.035 & 0.022 & \textbf{0.020} ($\SI{17}{\percent}$ PID costs) \\
        \bottomrule
    \end{tabular}
    \caption{Cost of PID, LQR, MPC, and our BP-MPC scheme for the circle and square references. Controllers with an asterisk use pre-identified parameters $\theta$.}
    \label{tab:tracking_errors}
\end{table}

The results confirm that our BP-MPC scheme improves control performance over iterations. The tracking error of the untrained MPC is three times smaller than that of the LQR and PID. Errors further reduce by a third using a pre-identified $\theta$. Finally, optimizing the MPC parameters $p$ using BP-MPC next to system-identification reduces tracking errors by an additional $\SI{10}{\percent}$. 

\subsection{Does System Identification Improve BP-MPC?}
\begin{figure}[!t]
    \centering
    \includegraphics{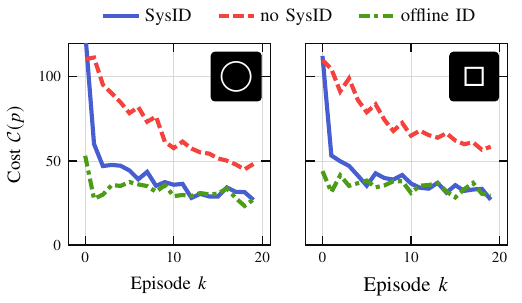}
    \caption{Closed-loop cost convergence trends for policy optimization framework across different system identification (SysID) modes for circle (left) and square (right) references.}
    \label{fig:PO_SI_comparison}
\end{figure}

We study the impact of system identification on BP-MPC performance by executing our scheme with and without system-identification as well as with a pre-identified system parameter. 
Figure \ref{fig:PO_SI_comparison} depicts the closed-loop costs of all experiments over training episodes for the circle and square references. 

\begin{figure*}[!tb]
    \centering
    \includegraphics{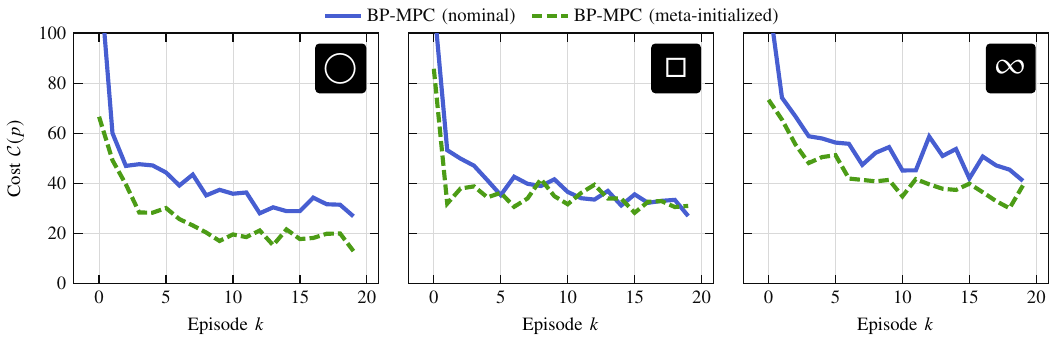}
    \caption{Closed-loop cost convergence trends for nominal and meta-initialized policy optimization across circle (left) and square (middle) training references, and an unseen figure-eight reference (right).}
    \label{fig:metalearning}
\end{figure*}

The system identification (i.e., step 5 in \cref{alg:exp:maml_no_maml}) improves the MPC prediction model and therefore prediction accuracy. 
It further reduces bias in the gradient estimates. 
This leads to two-times lower closed-loop costs of BP-MPC with system identification compared to BP-MPC without system identification (where step 5 is skipped altogether) at episode $20$ in Figure \ref{fig:PO_SI_comparison}. 
Notably, the cost-reduction can not only be attributed to the improved MPC prediction model thanks to system identification. 
Even with a pre-identified model (that is, for a scheme where step 5 is skipped but $\theta_r^0$ is accurate) BP-MPC reduces closed-loop costs by $\SI{33}{\percent}$ on the square reference, and by $\SI{44}{\percent}$ on the circle reference.

\subsection{Does Warm-Starting BP-MPC with MAML Improve the Performance on Unseen References?}

We now study whether our MAML algorithm obtains initial task specific parameters $p(\bar{p})$ which yield low tracking errors across different references, including unseen ones. We run \cref{alg:method:maml} using the circle and square references as training data and test the optimal parameter $\bar{p}$ without further training on the two training trajectories and additionally on an unseen figure-eight trajectory. The results are summarized in Fig.~\ref{fig:metalearning} and Tab.~\ref{tab:update_comparison}.

As expected, running the MPC with the optimal parameters for each respective reference always outperforms the MAML parameter $\bar{p}$. However, $\bar{p}$ yields strictly lower tracking errors on all references than parameters trained on different references. Further, even on the unseen figure-eight-reference, the MAML parameter $\bar{p}$ yields lower tracking errors across episodes than the untrained MPC parameter $p$.

This empirically confirms that implementing MAML yields initial parameters for BP-MPC with improved performance on unseen references. Note that this robust initialization does not render online updates redundant; learning optimal parameters using our BP-MPC scheme further reduced closed-loop costs for each reference by more than $\SI{20}{\percent}$.

\begin{table}[!tb]
    \centering
    \begin{tabular}{cccc}
        \toprule%
        & $\mathcal{C}_{\text{full}}(p^{\textit{circle}})$ & $\mathcal{C}_{\text{full}}(p_r(\bar{p}))$ & $\mathcal{C}_{\text{full}}(p^{\textit{square}})$ \\
        \midrule
        \textbf{Circle} & 32.0608 $\pm$ 6.6546 & 34.3113 $\pm$ 5.9981 &  42.4070 $\pm$ 7.0274 \\
        \textbf{Square} & 46.0763 $\pm$ 5.4495 & 43.6525 $\pm$ 5.3521 & 37.8114 $\pm$ 4.4533 \\
        \bottomrule
    \end{tabular}
    \caption{Closed-loop MPC cost for the circle and square reference for different parameters. $p^{\textit{circle}}$: parameters learned by BP-MPC for the circle-reference, $p^{\textit{square}}$: parameters learned by BP-MPC for the square task, $\bar{p}$: task-specific parameters learned using MAML BP-MPC (after a single gradient update). \label{tab:update_comparison}}
\end{table}
\section{Conclusion}
We presented a novel framework that integrates meta-learning and policy optimization to train highly adaptable Model Predictive Control (MPC) controllers. By adapting model-agnostic meta-learning (MAML) to a differentiable MPC framework, our approach successfully addresses the computational bottlenecks typically associated with closed-loop tuning. Furthermore, we incorporated online system identification into the end-to-end pipeline to continuously and concurrently refine both the underlying predictive models and the MPC hyperparameters.

Experimental validation on a hardware Ball-on-Plate testbed confirmed that combining system identification with gradient-based policy optimization improves closed-loop performance and accelerates convergence without adding computational costs. Furthermore, leveraging MAML provides robust initial parameters for rapid adaptation to completely unseen tasks with minimal overhead. Ultimately, this methodology enables the design of highly adaptable MPC controllers that maintain exceptional performance across varying tasks while mitigating the time and reset overheads common in real-world hardware implementations.

Future work will focus on strengthening the safety guarantees of the adaptation procedure and on developing accelerated algorithms through tailored Quasi-Newton approximations.

\section*{Acknowledgments}
We are grateful to André Warnecke for his support on on the BoP hardware. 
We are grateful to Jiaqi Yan for helpful discussions and to Alexander Kaspar for the promising earlier results motivating the current approach.

\bibliographystyle{IEEEtran}
\bibliography{IEEEabrv,bibliography}

\end{document}